# Non-Hermitian engineering of synthetic saturable absorbers

M.H. Teimourpour[1,2], A. Rahman[1,2], K. Srinivasan[3] and R. El-Ganainy[1,2,*]

[1]*Department of Physics, Michigan Technological University, Houghton, Michigan 49931, USA*

[2]*Henes Center for Quantum Phenomena, Michigan Technological University, Houghton, Michigan 49931, USA*

[3]*Center for Nanoscale Science and Technology, National Institute of Standards and Technology, Gaithersburg, Maryland 20899, USA*

**Abstract**

We introduce a new type of synthetic saturable absorbers based on quantum inspired photonic arrays whose linear light transport characteristics can be derived via bosonic algebra. We demonstrate that the interplay between optical Kerr nonlinearity, interference effects and non-Hermiticity through radiation loss leads to a nonlinear optical filtering response with two distinct regimes of small and large optical transmissions. More interestingly, we show that the boundary between these two regimes can be very sharp. The threshold optical intensity that marks this abrupt "phase transition" and its steepness can be engineered by varying the number of the guiding elements. The practical feasibility of these structures as well as their potential applications in laser systems and optical signal processing are also discussed.

PACS# (42.82.Et) Waveguides, couplers, and arrays, (42.65.-k) Nonlinear optics, (42.65.Wi) Nonlinear waveguides

[*]ganainy@mtu.edu

The introduction of parity-time reversal symmetry (PT) symmetry [1] in optics [2-6] has initiated an intense research program that aims at building new photonic devices with novel functionalities such as non-reciprocal devices [7,8], single mode PT lasers [9,10], supersymmetric laser arrays [11], Exceptional point-based sensors [12] as well as light sources based on non-Hermitian phase matching [13]. Still an area that remains largely unexplored (apart from the work in [10] and the recent results in [13]) is the engineering of the nonlinear non-Hermitian response function of photonic structures to achieve useful functionalities. In this work, we explore one such possibility and we provide a new photonic geometry that exhibit saturable absorber response with superior performance. As we will explain in details shortly, in our system the non-Hermiticity arises from first separating the undesired light components spatially and then introducing loss either through coupling to the continuum or optical absorption.

Saturable absorbers (SAs) are optical components that exhibit large/small optical absorption coefficients at low/high light intensity, respectively [14]. They are indispensable tools for a wide range of applications from mode-locking [15,16] and Q switching [15,16] to pulse shaping and stabilization [17,18] and noise suppression [19]. In general, saturable absorbers can be classified into two main categories based on their operation principle 1) devices that rely on engineered electronic band structures to achieve proper functionality, and 2) artificial saturable absorbers that exhibit a decrease in optical absorption at high intensities without having physical absorption saturation. Semiconductor SA mirrors [20], quantum dots based SA [21] and SA that utilize the optical/electronic properties of carbon nanotube [22] and graphene [23] are examples of the former. The latter includes Kerr lensing [24], fiber loops [25] and nonlinear waveguide arrays [26].

In particular, using uniform waveguide arrays as artificial saturable absorbers utilizes the interplay between Kerr nonlinearity, which is an intrinsic property of the material, and discrete diffraction which can be engineered by scaling the uniform coupling coefficients between the guiding channels [27]. This strategy has been used to investigate mode-locking using waveguide arrays [28-30]. In the past few years, different types of non-uniform waveguide arrays have been proposed and shown to exhibit interesting features. A particular structure that received considerable attention recently is the so called $J_x$ array-first conceived within the context of spin networks [31] - whose properties can be derived by using angular momentum algebra (hence the name $J_x$ after the angular momentum operator). Interestingly, it was shown later that the properties of $J_x$ arrays can be also investigated by using bosonic algebra. This connection which is anticipated from the Schwinger boson representation of angular momentum [32] has provided a tool for engineering quasi-one dimensional parity time (PT) symmetric structures that exhibit higher order exceptional points [33]. Among the intriguing characteristics of $J_x$ photonic arrays is coherent transport between ports that are related through mirror reflection around the array's axis of symmetry [34]. This feature was recently utilized in a theoretical proposal to build on-chip optical isolators and polarization beam splitters [35,36]. In addition, due to their ladder eigenvalue spectrum, revival effects have been also recently predicted in supersymmetric partners of $J_x$ arrays [37] despite the fact they do not possess any spatial symmetry.

To date, nonlinear light interactions inside $J_x$ arrays have not been investigated. In this letter, we consider the effect of Kerr nonlinearity on light propagation in $J_x$ arrays and we show that under the appropriate conditions, these structures can function as a synthetic digital saturable absorber, i.e. as a SA with an extremely sharp transition between the low and high transmission regimes as a function of the input optical intensity.

Figure 1 depicts a schematic illustration of a photonic $J_x$ waveguide array. Within the coupled mode formalism, continuous wave (CW) light propagation inside a $J_x$ array having a total number of guiding elements equal to $2N+1$ (with $N$ being integer or half integer number) is given by:

$$i\frac{dE_n}{dz} + \kappa(g_{n-1}E_{n-1} + g_n E_{n+1}) + \chi|E_n|^2 E_n = 0 \qquad (1)$$

In Eq.(1), $E_n$ is the electric field envelope amplitude in waveguide $n \in [-N, N]$, $z$ is the propagation distance while $\kappa$ is a scaling constant that characterizes the coupling between the different elements and the coupling profile is given by $g_n = \sqrt{(N+n)(N-n+1)}$, $g_{-N} = 0$ and $g_{N+1} = 0$ (see references [31,33,36] for a detailed derivation of the coupling profile $g_n$). Furthermore, $\chi$ the effective nonlinear Kerr coefficient and it is proportional to the material nonlinearity and the optical mode confinement factor [38]. Figure 1 depicts a schematic of a $J_x$ array made of seven waveguides ($N=3$). The input and output ports of the device are also indicated on the figure. The additional output ports serve as a means for introducing optical losses to the system by coupling the undesired light components in these channels to the radiation continuum. In practice, depending on the modal profile and phase mismatch between the waveguide modes and free space propagation, small fraction of this light will be reflected and transmitted back to the input channel. This feedback process might hamper the operation of the device. This can be overcome by extending the auxiliary ports beyond the device length, bending them away from the main port and introducing physical optical absorption on each channel through metal deposition. These design details will be explored in future studies.

We proceed by using the scaling $\xi = \kappa z$ and $E_n = \sqrt{P_o} a_n$, where $\xi$ and $a_n$ are the normalized distance

and field amplitude respectively while $P_o$ represents the input optical power, we finally obtain:

$$i\frac{da_n}{d\xi} + (g_{n-1}a_{n-1} + g_n a_{n+1}) + \alpha |a_n|^2 a_n = 0,$$ where, the nonlinear parameter is given by

$\alpha = \chi P_o / \kappa$. Under linear conditions, when $\alpha = 0$, light propagation in these arrays undergoes coherent transport where an optical beam launched in one port will exit from the mirror symmetric port with a unit power transmission [31,32] after a certain propagation distance. As a result, in this case, the transmission coefficient between input and output ports $T$ belonging to the same channel (leftmost waveguide in Fig.1) will be zero. When the input light intensity is increased, or equivalently by increasing $\alpha$, the transmission coefficient $T$ is expected to grow as the nonlinear self-trapping effects starts to overcome discrete diffraction [27]. These intuitive predictions are confirmed in Fig.2 (a) where we plot the values of $T$ as a function of the nonlinear coefficient $\alpha$ or equivalently the input light intensity for three different $J_x$ arrays made of three, seven and eleven guiding elements, respectively. In all arrangements, the input light is launched in the leftmost waveguide. The normalized length of each $J_x$ array is chosen to be $\xi = L \equiv \pi/2$ in order to produce exactly zero transmission when $\alpha = 0$ (see ref. [27] and references therein). Interestingly, we observe a very sharp transition between the low and high transmission states (off/on states) as a function of the nonlinear parameter $\alpha$. In addition, the location at which this transition takes place, $\alpha_{TH}$ defined as the threshold value of $\alpha$ at which the transmission starts to rise above say $5 \times 10^{-3}$ monotonically (alternatively one might chose other criteria for $\alpha_{TH}$ such as the point where the slope of the curve changes to a certain value), is a function of the waveguide number. For comparison, we also study the dynamics in uniform waveguide arrays. If the array is infinite, the uniform coupling $\kappa_u$ can be chosen to produce exact zero transmission under single excitation conditions after a propagation distance of $L$, i.e. it satisfies the relation $J_o(2\kappa_u L) = 0$, where $J_0(x)$ is the Bessel function of order zero and argument $x$ (see Ref. [27] for discussion and references therin for detailed derivation). A large but finite array can still exhibit near zero linear transmission similar to an infinite array if we maintain the above condition and launch light only in the central channel. Fig.2 (b) shows the transmission curves for uniform arrays made of three, seven and eleven waveguides when light is launched in the central element of the array. Clearly the transition between the two regimes in this latter case is less abrupt than in the case of $J_x$ array (the slope of the curve for uniform array with eleven waveguides is ~10% of that of the case of $J_x$ array having same number of channels). Finally, for completeness we also consider the case of edge coupling in a simple directional coupler case as well as the case of uniform arrays with three, seven and eleven channels as depicted in Fig.2 (c). In the directional coupler case, we observe smooth but not very abrupt (the slope of the curve is ~16% of that of the case of $J_x$ array having eleven waveguide channels) transition. For larger uniform arrays, we observe an interplay between sharp transitions and fast oscillations. Thus, based on the above simulations, we find that the nonlinear response of the $J_x$ structures provides a superior performance in the sense that it is very abrupt and free of any oscillatory behavior. These results might find applications in several different photonic systems as we will discuss later.

In order to illustrate the physics behind the observed abrupt transition between the off/on states (Fig.2 (a)), we investigate light transport dynamics in a $J_x$ photonic waveguide array made of seven guiding channels after a propagation distance $\xi = L \equiv \pi/2$ when light is launched in the left most channel. Under linear conditions, $\alpha = 0$, we obtain coherent perfect transfer as expected (Fig.3 (a)). As the nonlinear parameter increases, self-trapping effects take place and reduces the transmission to the rightmost port as shown in Fig.3 (b) for $\alpha = 10$. However, in this regime, the nonlinear response is not strong enough to steer appreciable optical energy to the leftmost channel from adjacent waveguides. As the input light intensity is further increased, localization effects becomes stronger and just at $\alpha = 11.3$, transmission through the output port (leftmost channel) starts to rise as illustrated in Fig.3 (c). Finally as shown in Fig.3 (d), at $\alpha = 12$ we observe strong spatial soliton effects where light remains mainly localized in the input waveguide and thus leads to a high transmission coefficient. This same behavior is observed for arrays made of different number of guiding elements with the difference that stronger nonlinearities (or higher light intensities) are needed to divert light from right to left when the number of the waveguides increases. This explains the behavior observed in Fig.2 (a) where the transmission curve shifts as a function of the number of the waveguides $M$. This last point is better illustrated in Fig.4 where we quantify the dependence of $\alpha_{TH}$ and the sharpness of the transitions - defined by the slope of the transmission curve in the transition region (*TR*) and denoted by $(dT/d\alpha)_{TR}$ and computed by using a linear curve fitting at the central part of the transition region- as a function of the number of waveguides. Interestingly, both curves can be fitted with quadratic functions with a maximum error of 5% and 7%, respectively. This excellent quadratic fitting hints at the possibility of semi-analytical solutions for the nonlinear dynamics and soliton effects that underlines the nonlinear optical response of these structures which we plan to investigate in details elsewhere.

Intuitively, one would also expect other ports in the array (except the central one) to function as synthetic saturable absorbers. However, as shown in Fig.5 for an array made of seven guiding channels, when light is launched and collected from port 2 or 3 (second leftmost waveguide or third leftmost one), interference effects become dominant in the transition region, giving rise to oscillatory transmission behavior before it eventually settles down. On the other hand, when light is launched in the central channel (waveguide number four in this example), the nonlinear response resembles that of a band-stop filter with a reduced transmission band between two high transmission regions at very low and very high intensities- a feature that might find applications in controlling nonlinear laser dynamics. The nonlinear wave propagation dynamics that leads to the behavior observed in Fig.5 are depicted in Fig.6 for the case of central port excitation for different values of the nonlinear parameter $\alpha$.

Now we briefly discuss several practical aspects of proposed synthetic saturable absorber structure: **(1)** *Experimental realization:* we first note that linear $J_x$ photonic arrays have been demonstrated in silica glass platforms [31]. Observing saturable absorption effects in these silica-based structures [31] is possible at relatively higher optical intensities [38]. Another attractive alternative is to use *AlGaAs* platforms which exhibit higher nonlinearities (almost three orders of magnitude higher than silica) [39] and is compatible with semiconductor laser technology- an advantage that might lead to integrating the laser device and the synthetic SA on the same chip to build a micro-scale mode-locked lasers. **(2)** *Nonlinear response scaling***:** in addition to the predicted sharp "phase" transition between the low and high transmission modes (a feature that cannot be achieved by using uniform arrays) another advantage offered by our proposed structure is the ability to tailor the nonlinear response of the system by scaling the coupling constants between the waveguide elements without the need to employ a new material system. In addition, being a waveguide array-based device, the SA proposed in this work is expected to enjoy a relatively large bandwidth (a feature that we plan to investigate in future work by using full wave analysis). **(3)** *Ultrafast applications***:** Another interesting feature of the proposed device is that, similar to the work in [28-30], the absorption saturation effects rely on Kerr nonlinearity which is an instantaneous effect that does not depend on any carrier relaxation mechanism and thus suitable for ultrafast applications. We note that our analysis in this work was based on CW light wave propagation. For ultrafast applications, one has to modify Eq.(1) to include temporal dispersion as well as additional nonlinear effects [38]. The details of these terms however are material and structure dependent and we will investigate them in details in conjunction with specific material systems and particular photonic designs in future publications.

In conclusion, we have proposed a new concept for building synthetic saturable absorbers based on photonic $J_x$ arrays. We have investigated the nonlinear response of these structures and we have studied their light transport dynamics. Our simulations uncovered an interesting behavior of sharp "phase transition" between two different transmission regimes: off/on states. This feature can be utilized to build on-chip mode-locked lasers as well as optical comparator devices that differentiate between zero and one states.

**Figure captions**

Fig.1. (Color online) Schematic of a nonlinear $J_x$ array made of seven waveguide elements. The coupling profile is symmetric around the central channel and values of the coupling coefficients are indicated. The array length $\xi = L \equiv \pi/2$ is chosen to provide coherent perfect transfer under linear condition. The Kerr nonlinear coefficient in these arrangements is a function of the material properties and the mode confinement inside the waveguide. In our proposed saturable absorber scheme, light is launched in and collected from in the leftmost port. The other additional ports serve as a mean to introduce optical loss either by direct radiation into free space or by introducing physical loss as explained in the text.

Fig.2. (Color online) (a) Transmission coefficient between the input and output ports as defined in Fig.1 as a function of the nonlinear parameters (or input light intensity) for $J_x$ arrays made of three, seven and eleven waveguides, respectively. (b) A comparison with the performance of a uniform array as described in the text. Evidently, $J_x$ arrays provides an abrupt transition between the off/on regimes -an effect that is not observed in uniform arrays. Note that the nonlinear response of the uniform array for $M=7$ and 11 is almost identical. The black arrows in (a) indicate roughly the location of $\alpha_{TH}$ on each curve. The transition region is indicated on the $M=11$ curve by the black and red arrows. (c) Nonlinear response under edge excitations in a simple directional coupler with coupling coefficient $\kappa_u = 1$ (curve C) and uniform arrays with seven waveguide channels when the coupling is $\kappa_u = 1$ (S$_1$), $\kappa_u = 2$ (S$_2$), and $\kappa_u = 3$ (S$_3$), respectively. In the directional coupler case, we observe smooth but not very abrupt (compared to the $J_x$ array) transition. For larger uniform arrays, we observe a smooth transition for small coupling coefficients,. As the coupling increase, the transition becomes sharper and is accompanied with fast oscillatory behavior.

Fig.3. (Color online) Light transport dynamics in a nonlinear $J_x$ array made of seven waveguides under different input light intensities as quantified by the value of the nonlinear parameter: (a) $\alpha = 0$, (b) $\alpha = 10$, (c) $\alpha = 11.3$ and (d) $\alpha = 12$. For a wide range of $\alpha$, the self-trapping effects are not strong enough to prevent light diffraction and forces the optical intensities to be confined in the same input channel. However at a certain light intensity threshold (defined in the text by $\alpha_{TH}$), strong spatial soliton effects emerge, leading to a sudden and sharp increase in the transmission coefficient.

Fig.4. (Color online) The dependence of the threshold intensity characterized by $\alpha_{TH}$ and transition steepness defined by $(dT/d\alpha)_{\alpha_{TH}}$ on the array size. Both curves follow a quadratic dependence within maximum errors of 5% and 7%, respectively.

Fig.5. (Color online) Transmission dynamics in a $J_x$ array made of seven waveguides as a function of the input light intensity when light is launched and collected from the same channel for waveguides number two, three, and four respectively after a normalized propagation distance of $\xi = L \equiv \pi/2$. Note that interference effects play an important role in these cases, leading to oscillatory behavior. When light is launched in the central channel, a low transmission band

(defined by intensity range rather than frequency range) exists between high transmission bands- a feature that might find applications in laser systems.

Fig.6. (Color online) Nonlinear wave propagation in the $J_x$ array studied in Fig.5 when light is excited at the central waveguide for $\alpha = 0, 9, 12, 14.5, 16.5$ and $18$, respectively. From the plots, we can see that the interplay between nonlinear self-localization effects, reflection from the array edges and interference lead to the oscillatory transmission behavior observed in Fig.5.

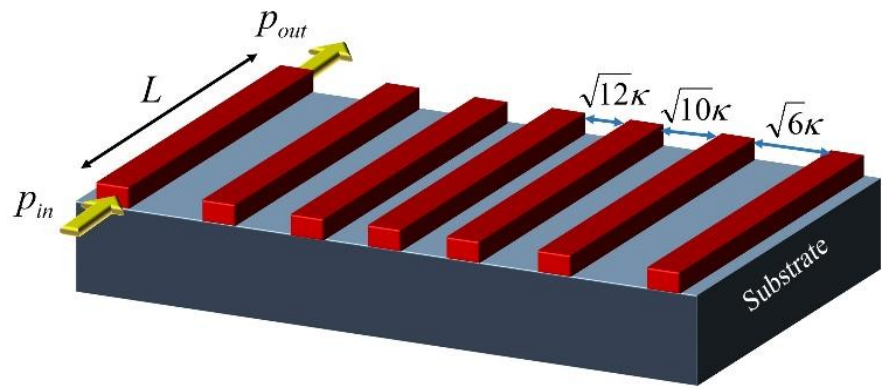

Fig.1

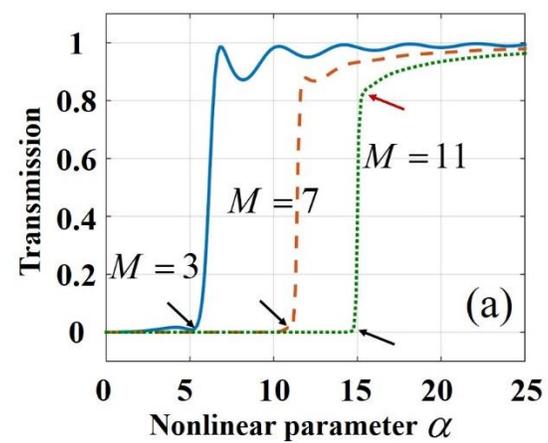

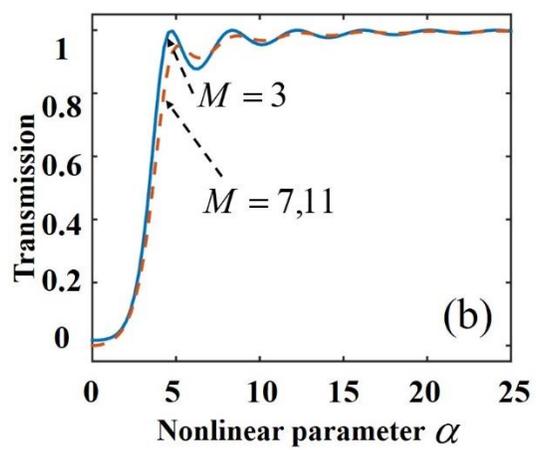

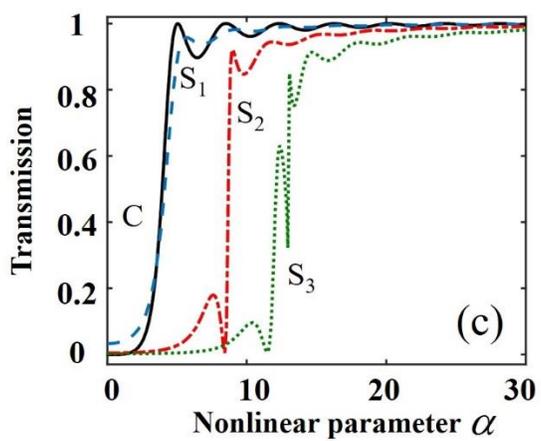

Fig.2

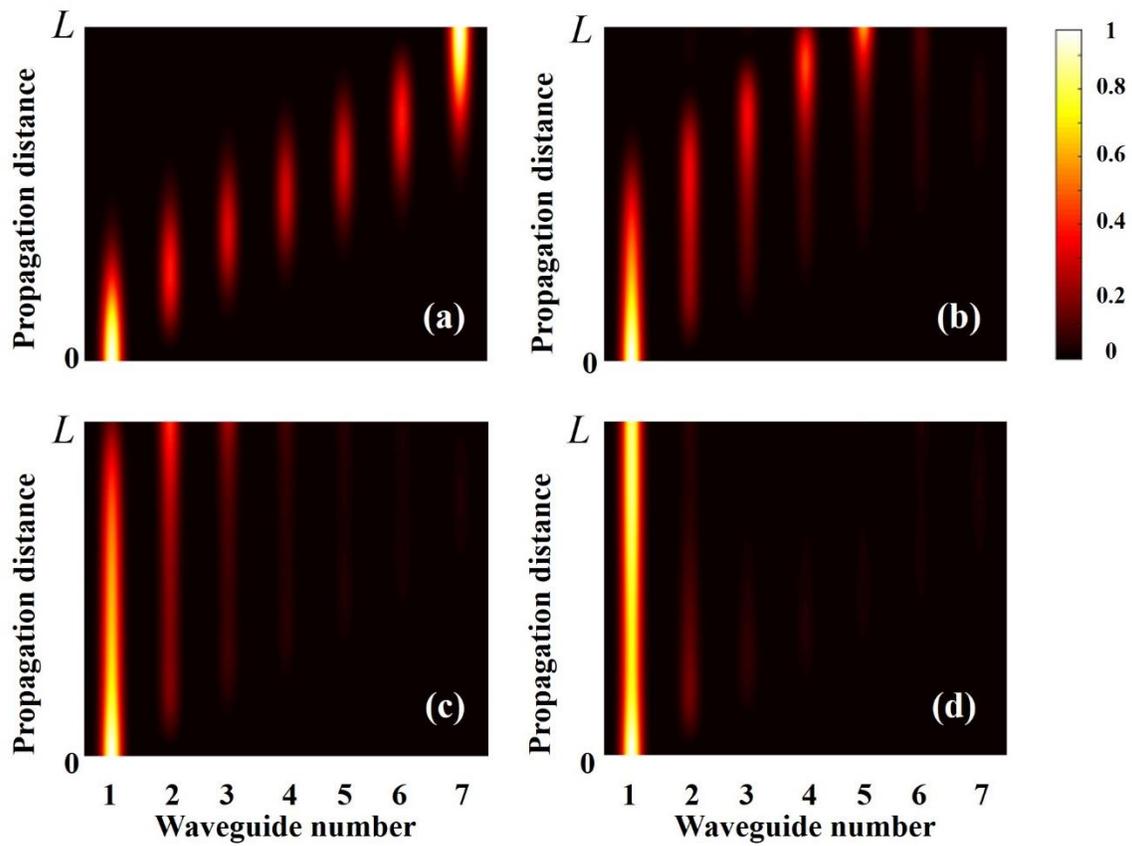

Fig.3

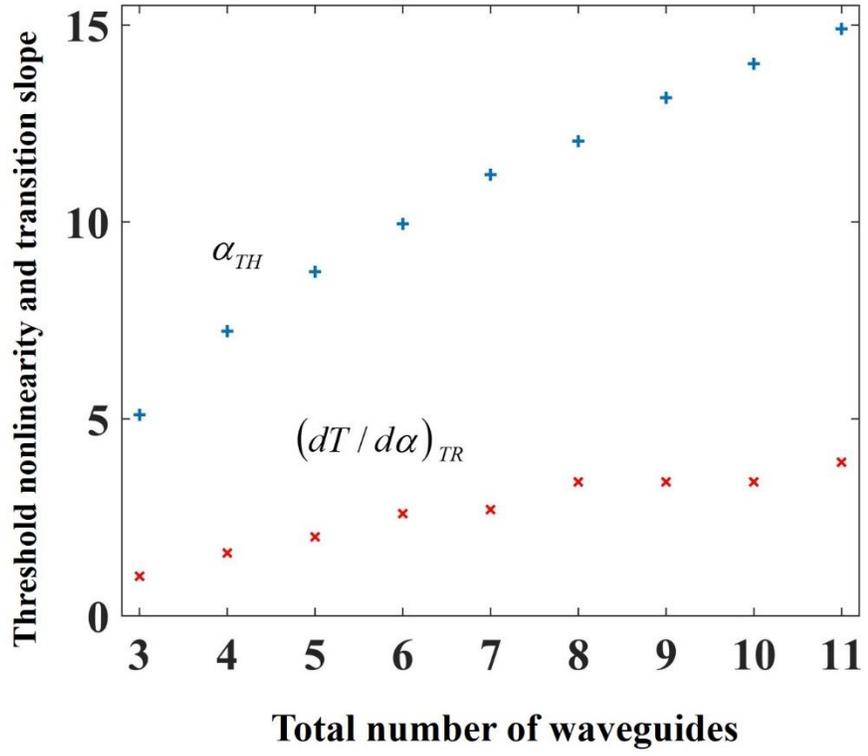

Fig.4

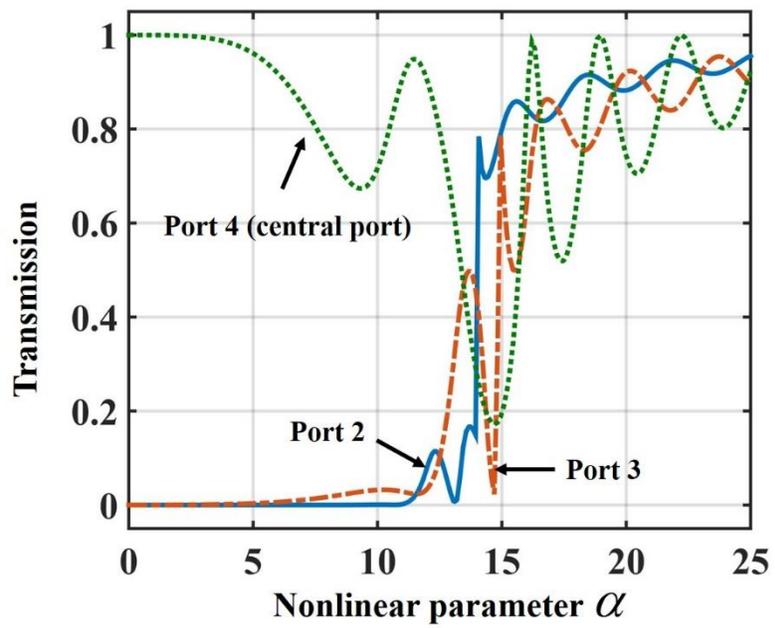

Fig.5

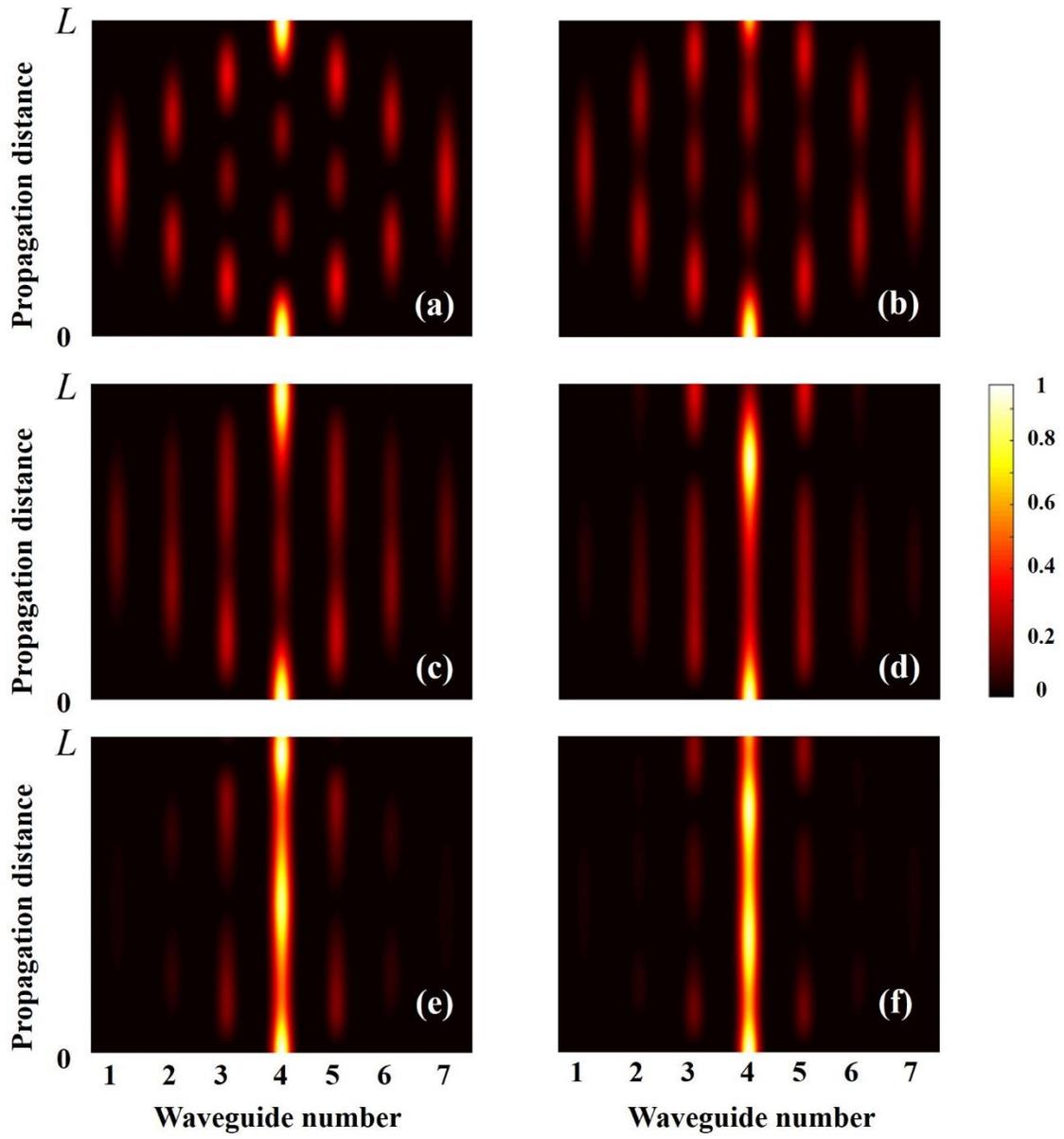

Fig.6